\documentclass{article}

\usepackage{textcomp}
\usepackage{pifont}
\usepackage{amsmath}   
\usepackage{subfigure} 
\usepackage{icrctc07}

\title{A Geant4 based engineering tool for Fresnel lenses}
\shorttitle{A Geant4 based engineering tool for Fresnel lenses}
\authors{Jo\~ao Costa$^{1}$, M\'ario Pimenta$^{2}$, Bernardo Tom\'e$^{2}$.}
\shortauthors{J. Costa and et al}
\afiliations{
$^1$ 
Laboratoire de l'Acc\'el\'erateur Lin\'eaire, 
IN2P3-CNRS et Universit\'e Paris-Sud 11,
Centre Scientifique d'Orsay, B. P. 34, 91898 Orsay, Cedex, France. 
\\
$^2$ Laborat\'orio de Instrumenta\c{c}\~ao e F\'{\i}sica 
Experimental de Part\'{\i}culas, Av. Elias Garcia, 14--1, 1000-149 Lisboa, Portugal.
}

\email{bernardo@lip.pt}

\abstract{
Geant4 is a Monte Carlo radiation transport toolkit that is becoming  a tool of generalized application 
in areas such as high-energy physics, nuclear physics, astroparticle physics, or medical physics.
Geant4 provides an optical physics process category,
allowing the simulation of the production and propagation of light.  
Its capabilities are well tailored for the simulation of optics systems namely in cosmic-rays experiments 
based in the detection of Cherenkov and fluorescence light. 
The use of Geant4 as an engineering tool for the optics design and simulation of Fresnel lens systems is 
discussed through a specific example. 
}

\begin{document}
\maketitle

\section{Introduction}
The use of Fresnel lenses has been discussed in the last few years in the 
context of cosmic-rays experiments based on the detection of the Cherenkov and fluorescence light
produced by extensive air showers~\cite{EUSO,ULTRA,GAW,Lamb}. 
%
In the specific case of Cherenkov telescopes, the field of view (FOV) can be considerably enlarged by
employing Fresnel lens based optics. 
In fact, the FOV of exhisting and planned Cherenkov telescopes, using reflective optics, 
is limited to few degrees due to  the degradation of the imaging quality  
for off-axis angles and to the need of using small cameras at the focal plane to limit the mirror 
obscuration.   
To cover a large area of the sky, a possible alternative is to use refractive optics 
with the camera located behind the lens. 
Fresnel lenses are a viable solution due to their small thickness and lightness.


In this paper a simulation tool for Fresnel lenses using Geant4 is presented. 
It is shown how it can be explored  to perform realistic simulations and optimizations 
of optics systems using these lenses.


\section{The Geant4 toolkit}
Geant4 is a toolkit to simulate the particle transport and interactions in matter,
with tracking capabilities in 3D geometries of arbitrary complexity.
%
It includes an extensive set of electromagnetic, hadronic and 
optics physics processes~\cite{Geant4,Pia}. 
Optical photons are generated through scintillation, Cherenkov and transition radiation or can be 
explicitly emitted by a user defined light source. 
The tracking of optical photons includes refraction and reflection at medium boundaries, 
Rayleigh scattering and bulk absorption.  

\section{Lens geometry implementation}  

A Fresnel lens is constructed from a given lens surface by decomposing it into small pieces. 
Each  piece of the original lens surface is translated to the plano side of the lens,
thus defining a groove of the Fresnel lens.

%
%
In the present simulation the lens grooves have a conical shape,
each being a frustum of a cone with a cross-section represented by a straight line.
The slopes of these lines are obtained from the sagita equation of the original lens surface.
The concentric grooves have the same radial span --  constant width grooves. 
The geometry of a Fresnel lens is defined in Geant4 using a parameterized replication of G4Cons 
volumes. 
A  lens can be described either as a monolithic lens or as a generic lens petal, 
covering a limited azimuthal range. A generic lens petal can be rotated around a specified axis.




\section{Simulation of a specific lens}
As a working example a 1.7~m diameter plano-convex Fresnel lens consisting of a central monolithic lens and two
concentric rings of lens petals was simulated. 
This example was inspired by the innovative approach proposed by the GAW collaboration.


The lens profile, an aspheric even surface, was based on an optimization at $\lambda = 320$~nm
using the OSLO\cite{OSLO} optics software. 
%
%
%
The focal length is 2025~mm and 
the lens material is a typical ultraviolet (UV) transmitting acrylic.
Two lenses were simulated, with groove densities of 2~grooves/mm and 0.333~grooves/mm.

\subsection{Analysis of the optics}
In the lens design and optimization several parameters must be taken into consideration, 
namely the magnitude of the aberration, characterized by the point spread function (PSF), 
the lens transmittance and the telescope FOV. 
For the purpose of characterizing the present lens, some of its optimization parameters were varied: 
the density of  the grooves, the thickness of the lens, and the distance from the lens to 
the detection plane. 
Some of the results obtained in this context are shown hereafter with the main purpose of illustrating the 
capabilities of this simulation tool. A complete optimization of a detector and the systematic 
characterization of its performance is out of the scope of this paper.

\begin{figure}
	\centering
		\includegraphics[width=0.925\linewidth]{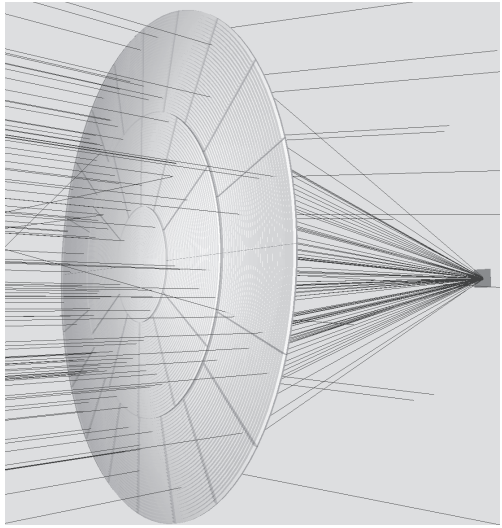}
	\caption{Visualization of the simulated Fresnel lens, illustrating the focusing of on-axis light incident from the left.}
	\label{fig:zoomlensfocus}
\end{figure}

\begin{figure}
	\centering
	\includegraphics[width=0.80\linewidth,angle=-90]{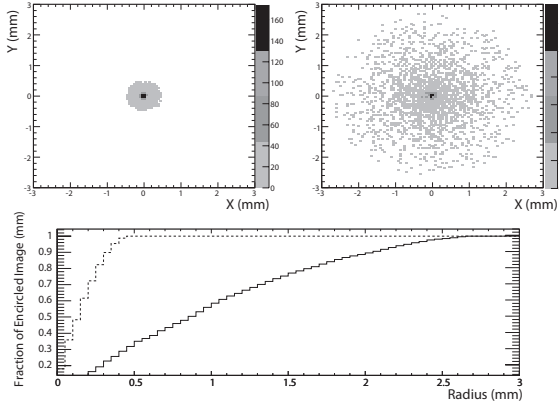}
	\caption{PSF for on-axis incidence of monochromatic light ($\lambda = 320 \mathrm{nm}$).
Upper left:  lens with groove density of 2/mm. Upper right: lens with groove density of 0.333/mm. 
Bottom: Encircled image {\it vs} radius for lenses with groove density of 2/mm (- - -) and 
 0.333/mm (---).}
	\label{fig:spotsacum}
\end{figure}

As a first example, the dependence of the PSF with the groove density 
was studied for UV light ($\lambda=$320~nm) incident parallel to the lens optical axis 
($\theta = 0$\textdegree), as shown in Fig.~\ref{fig:zoomlensfocus}.
The spot images at the nominal focal distance and the corresponding encircled image, 
the fraction of photons collected inside a circle of variable radius, 
are shown in Fig.~\ref{fig:spotsacum}.

The focusing was quantified by the radius of the circle at the focal plane containing 
90\% of the detected light (R$_{90}$). 
In the following studies a light beam incident on-axis was considered, yielding a PSF which is 
symmetric about the lens axis. However, for large incidence angles coma aberration distorts the PSF 
and the definition of the spot size should be modified.    

\begin{figure}
	\centering
		\includegraphics[width=0.985\linewidth]{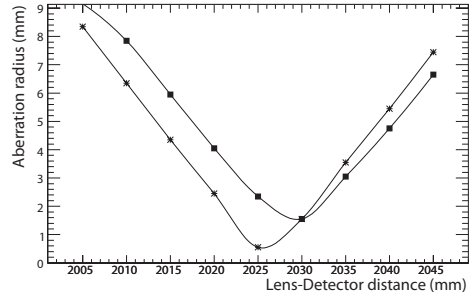}
	\caption{Aberration radius (R$_{90}$) as a function of the distance from the 
lens to the detector for on-axis incidence of monochromatic light ($\lambda = 320 \mathrm{nm}$). Results are shown for lenses with 2~grooves/mm (\ding{83}) and 0.333~grooves/mm (\ding{110}); the lines join the points.}
	\label{fig:optimizacoes}
\end{figure}

The dependence of R$_{90}$ with the focal distance is shown in Fig.~\ref{fig:optimizacoes}, 
for the two simulated lenses. 
For the lens with the smaller grooves the optimum distance $d_{min}$,
the distance that minimizes  R$_{90}$, is found to be close to the
nominal focal length. In the case of the lens with smaller groove density 
the aberration can still be reduced by about 60\% through the displacement of the focal plane by 5~mm 
from the nominal position.
In both cases the spot size at  $d_{min}$  follows approximately the width 
of the grooves of the corresponding lens.
This aberration is due to the  flat-surface groove facets: 
parallel photons arriving at a given groove are all refracted 
by the same angle, exiting the lens in a parallel beam, without undergoing any focusing. 
This intrinsic aberration, proportional to the size of the groove, is thus a characteristic 
of Fresnel lenses featuring flat surface grooves.

In a real cosmic ray experiment, a Cherenkov telescope will operate in an extended wavelength range
and the lens optimization should take into account effects such as the chromatic aberration. 
The Cherenkov spectrum can be incorporated  in the optimization studies thanks to the capabilities of Geant4.  
Effects such as the wavelength dependence of the light 
scattering in the atmosphere, of the light absorption in the lens material,
or of the spectral sensitivity of the photo-detectors can also be easily described with Geant4. 

\begin{figure}[htb]
	\centering
		\includegraphics[width=0.985\linewidth]{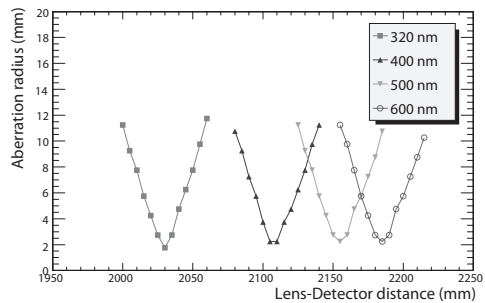}
	\caption{Aberration radius (R$_{90}$) as a function of the distance from the lens to the detector for different wavelengths.}
	\label{fig:optimizacao_vs_wavel}
\end{figure}

Figure~\ref{fig:optimizacao_vs_wavel} shows R$_{90}$  as a function of the focal distance 
for monochromatic light at several wavelengths. The size of the smallest PSF is approximately
the same for all  wavelengths, although at different  $d_{min}$. 
Since the refractive index of the lens decreases with the wavelength, $d_{min}$
increases with increasing wavelength, ranging from about 2030~mm for $\lambda = 320$~nm to 
2185~mm for $\lambda = 600$~nm. 

%

\begin{figure}
	\centering
         \includegraphics[width=0.985\linewidth]{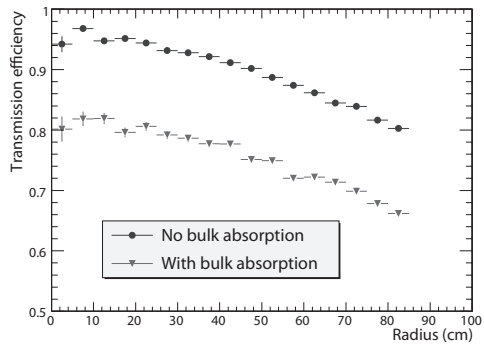}
\caption{Transmission of light with distance from center of the lens.}
\label{fig:transmissionwithradius}
\end{figure}
\begin{figure}
	\centering
		\includegraphics[width=0.475\linewidth,angle=-90,viewport=145 50 480 750,clip]{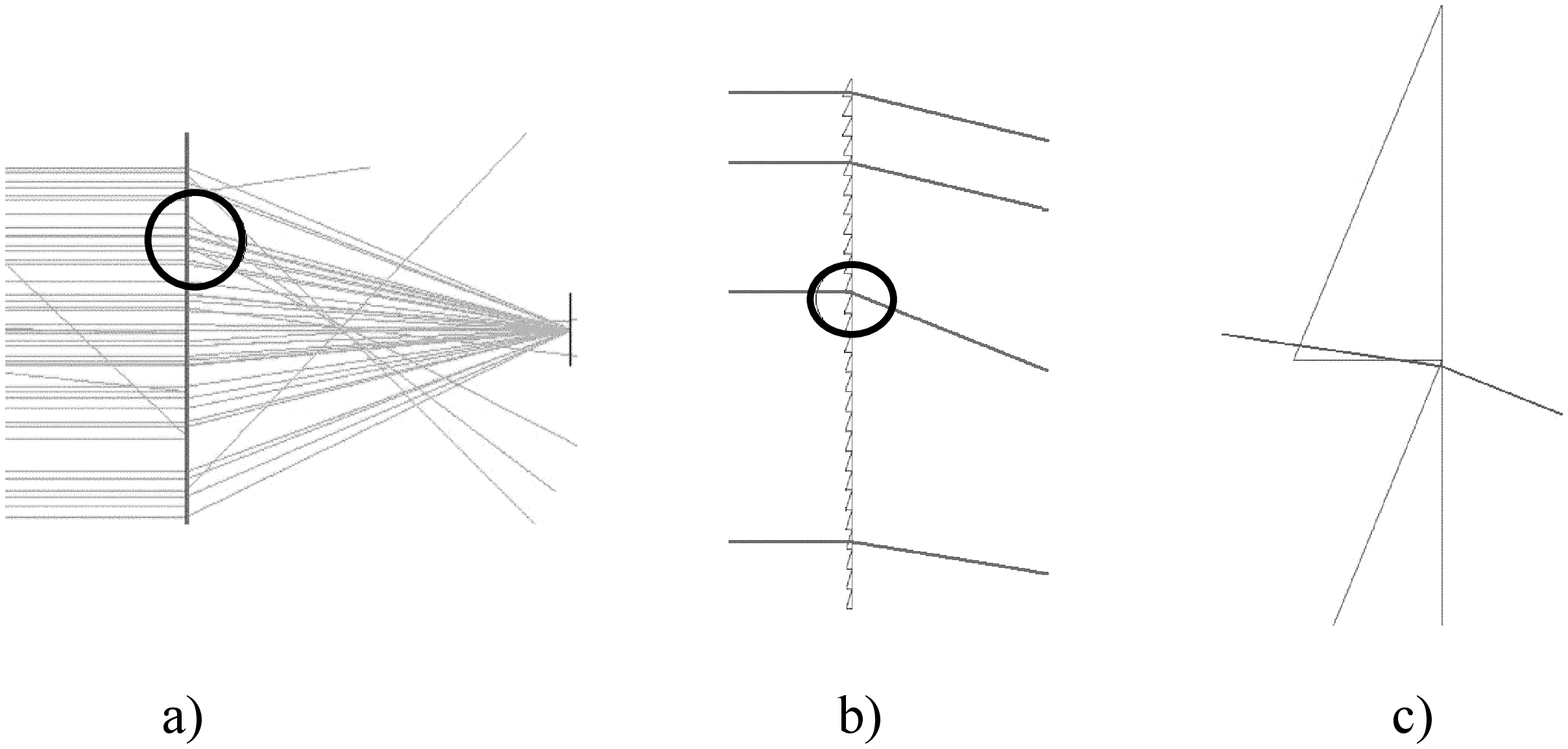}
	\caption{Example of multiple refraction of photons in the lens grooves. a) 
Some light rays  exit the lens out of focus. The encircled region in a) is shown in greater detail in b) and c), illustrating how photons impinging close to the edge of one groove can be refracted to the next groove.}
	\label{fig:picture1}
\end{figure}

The lens transmittance was studied as a function of the angle of incidence 
and of the distance from the center of the lens.  
Figure~\ref{fig:transmissionwithradius} shows the fraction of detected light as a function
of the distance of the incident rays from the lens axis, for $\theta = 0$\textdegree. 
Two cases are shown: a lens with infinite absorption length and a lens with the typical absorption spectrum of the UV 
transmitting acrylic. 
In both cases the light loss increases by about 20\% towards the edge of the lens.
While the bulk absorption, of the order of 10\% - 15\%, is approximately constant with the radius,
there is an additional effect reducing the light transmission which increases with the radius. 
As illustrated in Fig.~\ref{fig:picture1} this is due to the multiple refraction 
of photons which exit one groove near its border and cross an adjacent one, 
thereby changing their direction out of focus. 
As observed, this effect increases towards the lens periphery since, 
for constant width grooves, the groove height increases with the radius. 

Besides the type of analysis described here, which can be carried out in standalone mode,
it is possible to integrate the lens simulation in a complete Geant4-based detector simulation, 
including detector components such as light guides and  photomultipliers. 
Such simulation can also be interfaced 
with an external air-shower generator or a readout electronics and signal digitization module. 
These possibilities are particularly useful for the simulation of fluorescence and Cherenkov 
telescopes for present and future cosmic-ray experiments.

\section{Summary}

In this paper  an engineering tool for the optics design and simulation of Fresnel lenses using the Geant4 
toolkit was presented. Performance and optimization studies of a specific lens design were described. 

\section{Acknowledgements}
The authors from LIP acknowledge the support of Funda\c{c}\~ao para a Ci\^encia e Tecnologia, Portugal.

\bibliography{libros}
\bibliographystyle{plain}

\end{document}